\documentclass[aps,prb,superscriptaddress,reprint]{revtex4-1}

\usepackage{graphicx}
\usepackage{amsmath}
\usepackage[colorlinks]{hyperref}
\hypersetup{
   pdftitle={Resonant and off-resonant microwave signal manipulation in coupled superconducting resonators},
   pdfauthor={Mathieu Pierre, Sankar Raman Sathyamoorthy, Ida-Maria Svensson, G\"{o}ran Johansson, and Per Delsing},
   pdfkeywords={circuit-QED, superconductivity, microwave circuits, superconducting resonator, tunable coupling},
   urlcolor=blue,
   citecolor=blue,
   linkcolor=blue
   }

\begin{document}
\title{Resonant and off-resonant microwave signal manipulation in coupled superconducting resonators}
\author{Mathieu \surname{Pierre}}
\email[]{mathieu.pierre@lncmi.cnrs.fr}
\affiliation{Department of Microtechnology and Nanoscience (MC2), Chalmers University of Technology, SE-412\,96 G\"{o}teborg, Sweden}
\affiliation{Laboratoire National des Champs Magn\'{e}tiques Intenses (LNCMI), Universit\'{e} de Toulouse, INSA, CNRS UPR 3228, EMFL, FR-31400 Toulouse, France}
\author{Sankar Raman \surname{Sathyamoorthy}}
\author{Ida-Maria \surname{Svensson}}
\author{G\"{o}ran \surname{Johansson}}
\author{Per \surname{Delsing}}
\email[]{per.delsing@chalmers.se}
\affiliation{Department of Microtechnology and Nanoscience (MC2), Chalmers University of Technology, SE-412\,96 G\"{o}teborg, Sweden}

\begin{abstract}
 We present an experimental demonstration as well as a theoretical model of an integrated circuit designed for the manipulation of a microwave  field down to the single-photon level. 
 The device is made of a superconducting resonator coupled to a transmission line via a second frequency-tunable resonator. The tunable resonator can be used as a tunable coupler between the fixed resonator and the transmission line. 
 Moreover, the manipulation of the microwave field between the two resonators is possible. In particular, we demonstrate the swapping of the field from one resonator to the other by pulsing the frequency detuning between the two resonators.
 The behavior of the system, which determines how the device can be operated, is analyzed as a function of one key parameter of the system, the damping ratio of the coupled resonators. We show a good agreement between experiments and simulations, realized by solving a set of coupled differential equations. 
 \bigskip

\end{abstract}

\maketitle

In quantum technology the interaction between quantum states of light and various degrees of freedom of matter can be controlled in a variety of systems.
Among them, macroscopic superconducting circuits cooled to millikelvin temperatures are developing as a platform to manipulate microwave photons and artificial atoms. They are easy to engineer because they are integrated electrical circuits.
This forms the field of circuit quantum electrodynamics (circuit-QED) \cite{wallraff2004,blais2004}.

Using electrical circuits for building quantum systems allows for a precise design of Hamiltonian parameters within a wide range \cite{schoelkopf2008}. Furthermore, some parameters can also be made tunable \textit{in situ}, for instance, the resonance frequencies of resonators and the transition frequency of artificial atoms, also known as quantum bits \cite{sandberg2008,palacios2008,wallquist2006,koch2007}.

It is also essential for many experiments and applications to have tunable couplings, or equivalently, lifetimes or linewidths.
Tunable couplings have already been demonstrated between qubits \cite{hime2006,niskanen2007,ploeg2007,fay2008,bialczak2011,chen2014},
between qubits and resonators \cite{allman2010,allman2014,whittaker2014,hoffman2011,srinivasan2014}, 
and between resonators \cite{tian2008,baust2015,wulschner2015}.

In this work we focus on the tunable coupling between a resonator and a transmission line.
This function is required in several types of applications. First, in quantum communication \cite{kimble2008}, it is envisioned that ``flying'' qubits are sent over long distances in the form of photons \cite{eichler2011} propagating between nodes acting as quantum memories and processors. These nodes could be implemented as microwave resonators coupled to qubits or other types of quantum systems. It has been shown that the transfer efficiency can be increased if one can adjust the couplings at both ends of the transmission chain \cite{korotkov2011,sete2015}. Adjusting the coupling between the transmission line and the terminating resonator to the temporal and spectral properties of the incoming wave packet can result in full absorption \cite{wenner2014}, which can be viewed as an impedance matching condition for the resonator \cite{afzelius2013}.  
Inversely, a resonator with tunable coupling can also be used to emit microwave photons contained in an arbitrary wave packet. This has only been achieved  with more complex schemes so far \cite{pechal2014}.
Furthermore, it is becoming possible to simulate complex quantum systems, such as many-body states of condensed matter, using arrays of superconducting resonators and qubits. For this purpose, tunable couplings are essential to implement arbitrary Hamiltonians. Even more interestingly, dynamic processes can be studied if the couplings can be tuned fast enough, on the timescale of the processes under study.

Since resonators are either capacitively or inductively coupled to transmission lines, a first approach to make the coupling tunable is to use a tunable circuit element, such as a tunable inductance, for instance a superconducting quantum interference device (SQUID) \cite{yin2013}. To allow for more complex manipulations of the microwave signals, a second approach is based on a dual resonator architecture. A high quality factor resonator, dedicated to the storage of microwave radiation, which can be viewed as a quantum node, is connected to a transmission line via a low quality factor resonator. This low-Q resonator permits the fast transfer, storage or retrieval, of the quantum information encoded in the microwave radiation. 
It has already been shown how parametric processes can be used for the coherent manipulation of the microwave signals, either by coupling the two resonators with a Josephson ring modulator \cite{flurin2015}, a flux-driven Josephson junction circuit \cite{sirois2015}, or with a superconducting qubit \cite{pfaff2017}. 
In our work, we use a similar dual resonator architecture, but our approach for the coherent control is different. We made the low-Q resonator frequency-tunable, and the resonators are simply capacitively coupled.

In a previous article, we demonstrated the storage of microwaves in a superconducting resonator by switching on and off this tunable coupler \cite{pierre2014}.
We showed that microwaves can be released from the storage resonator through the frequency-tunable low-Q coupling resonator at a varying rate.
We presented a sample that was engineered to show a high on/off coupling ratio. 
The goal of the current article is to extend this work by presenting a generic model for this coupled-resonator circuit, valid in a large range of parameters and supported by experimental data in good agreement with the theory.
We show that the behavior of each sample is governed by a single parameter, a ratio between coupling rates, which corresponds to the damping ratio of the coupled resonator system. We present an experimental comparison of two samples operating in two distinct regimes. One of the sample corresponds to the results already presented in our previous work \cite{pierre2014}. It is optimized for direct addressing of the storage resonator, which is done in the off-resonant coupling of the two resonators. For the second sample, we show that the storage resonator can be addressed through a swapping procedure exploiting the resonant coupling of the two resonators.

\section{System and model} 

\subsection{The measured system}

\begin{figure}[t]
\begin{center}
\includegraphics[width=1.0\columnwidth]{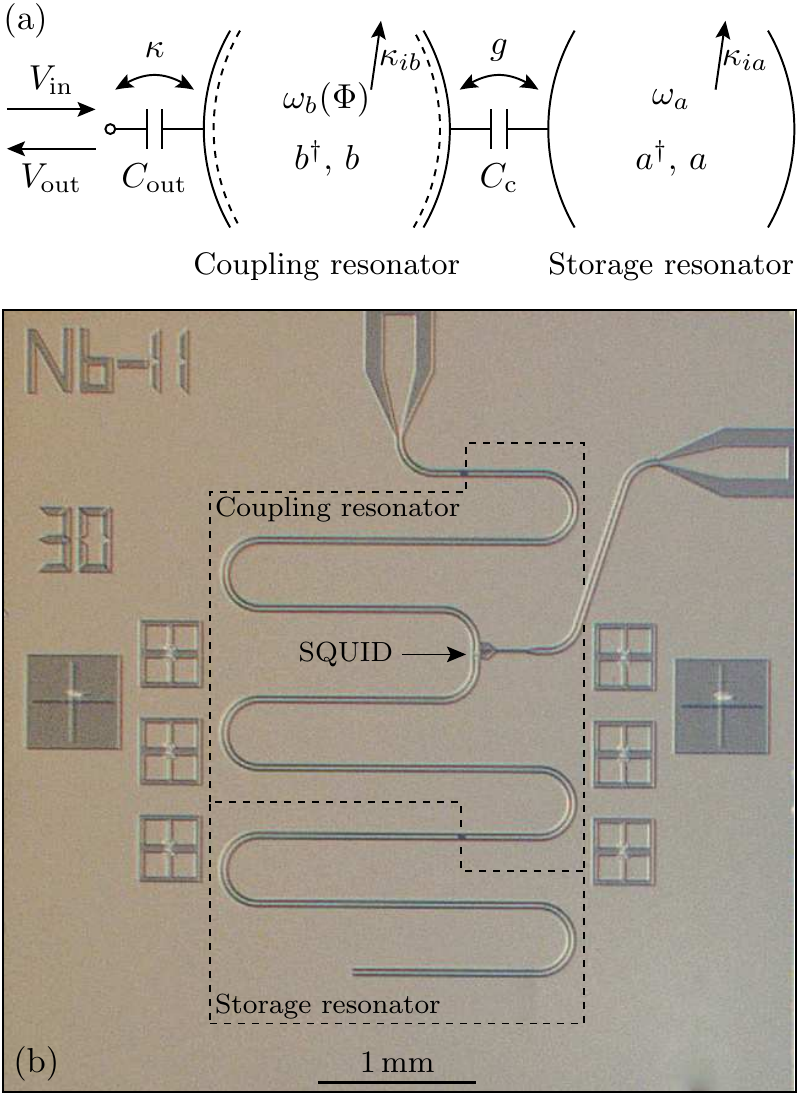}
\caption{(a) Model of the system under study. 
A storage resonator with frequency $\omega_a/(2\pi)$ is coupled to a transmission line via a coupling resonator with tunable frequency  $\omega_b/(2\pi)$.
(b) Optical microscope image of the corresponding superconducting integrated circuit (sample I).
} 
\label{fig0}
\end{center}
\end{figure}

The system under study is composed of two microwave resonators (see Fig.~\ref{fig0}). The resonators are coupled through a coupling capacitance $C_c$, permitting the transfer of energy between them.  
One of the resonators features a tunable resonance frequency. This allows to control the energy exchange between the two resonators, by changing their detuning.
The frequency tunability is based on a superconducting quantum interference device (SQUID). It behaves as a tunable, nonlinear, and nondissipative inductance embedded in the resonator \cite{sandberg2008,palacios2008}.

The tunable resonator has been engineered so that its range of reachable resonance frequency crosses the resonance frequency of the second resonator, which is constant. 
In addition, the tunable resonator is also coupled to a transmission line, which allows us to excite the system and probe it through microwave reflectometry. 
It will therefore be referred to as the coupling resonator, or resonator B. 
The other resonator contains no SQUID and thus has a fixed resonance frequency and a long lifetime. 
It is therefore suitable for microwave storage for instance \cite{pierre2014}, and will be referred to as the storage resonator, or resonator A.

\subsection{Theoretical model}

The theoretical model of the system is depicted in Fig.~\ref{fig0}(a).
In the rotating wave approximation, valid because the coupling rate $g$ between the resonators is much smaller than the resonator resonance frequencies $\omega_a$ and $\omega_b$, the Hamiltonian of the coupled resonators is 
\begin{align}
 \hat{H} = &\hbar \omega_a \hat{a}^\dagger \hat{a} + \hbar \omega_b \hat{b}^\dagger \hat{b} + \hbar g (\hat{a}\hat{b}^\dagger + \hat{a}^\dagger \hat{b}) \nonumber \\ &+ i\hbar \sqrt{\kappa}\left(   V_\text{in}^*(t) \hat{b} - V_\text{in}(t) \hat{b}^\dagger \right),
\end{align}
where $\hat{a}$ and $\hat{b}$ are the field ladder operators for resonators A and B, respectively, and $V_\text{in}(t)$ the input field driving the system.
Note that the Hamiltonian may be time dependent, as, in addition to the time-dependent drive, the resonance frequency of resonator B $\omega_b$ can be rapidly tuned in the experiment.

The coupling resonator is capacitively coupled to a transmission line, which makes the system open and dissipative. In addition, both resonators have finite intrinsic lifetimes, $1/\kappa_{ia}$ and $1/\kappa_{ib}$.
To describe the evolution of the quantum state of the system,
we use the Lindblad master equation \cite{lindblad1976,gardiner2004,breuer2002}, which gives the time evolution of the density matrix $\rho = \rho_a \otimes \rho_b$:
\begin{equation}
  \dot{\rho} = - \frac{i}{\hbar} \left[ \hat{H}, \rho \right] + \kappa \mathcal{D} [\hat{b}]\rho + \kappa_{ia} \mathcal{D} [\hat{a}]\rho + \kappa_{ib} \mathcal{D} [\hat{b}]\rho, \label{eq:master}
\end{equation}
where $\mathcal{D}$ denotes the Lindblad superoperator, defined as $\mathcal{D}[\hat{x}]\rho = \hat{x} \rho \hat{x}^\dagger - \frac{1}{2} \left\lbrace \hat{x}^\dagger \hat{x}, \rho \right\rbrace$.
Solving this equation gives the time evolution of the average photon number in each resonator, which cannot be measured directly in the experiment. For instance, for the storage resonator,
\begin{equation}
 \langle n_a \rangle = \langle \hat{a}^\dagger \hat{a} \rangle = Tr(\hat{a}^\dagger \hat{a} \rho).
\end{equation}

The classical response  of the system to the input field $V_\text{in}$ is given by the equations of motion for the expectation values of the resonator fields $A = \langle \hat{a} \rangle$ and $B = \langle \hat{b} \rangle$. They are derived using $ \dot{A} = Tr( \hat{a}\dot{\rho} )$ and $ \dot{B}  = Tr( \hat{b}\dot{\rho} )$.
\begin{align}
 \frac{dA}{dt} &= -i\omega_a A -igB - \frac{\kappa_{ia}}{2} A, \label{eq:motion1} \\
 \frac{dB}{dt} &= -i\omega_b B -igA - \frac{\kappa}{2} B  - \frac{\kappa_{ib}}{2} B - \sqrt{\kappa} V_\text{in}.   \label{eq:motion2}
\end{align}
The output voltage, which can be measured on the transmission line, is computed using the input-output relation
\begin{equation}
 V_\text{out} = V_\text{in} + \sqrt{\kappa} B.    \label{eq:in-out}
\end{equation}

In practice, the system is driven at an angular frequency $\omega_d$ close to the resonator resonance frequencies. It is therefore relevant to study its dynamics in a rotating frame. Natural choices for the rotating frame reference frequency are, for instance, the resonance frequency of the storage resonator $\omega_a$, which is constant, or the frequency of the drive field.
Redefining the resonator and output fields with respect to a reference frequency $\omega_\text{ref}$ by taking $A = ae^{-i\omega_\text{ref}t}$, $B = be^{-i\omega_\text{ref}t}$ and $V_\text{out} = v_\text{out}e^{-i\omega_\text{ref}t}$, and writing the driving field with respect to the driving frequency $V_\text{in} = v_\text{in} e^{-i\omega_dt}$, the equations of motion become
\begin{align}
 \frac{da}{dt} &= -i(\omega_a-\omega_\text{ref}) a -igb - \frac{\kappa_{ia}}{2} a, \label{eq:motion3} \\
 \frac{db}{dt} &= -i(\omega_b-\omega_\text{ref}) b -iga - \frac{\kappa}{2} b  - \frac{\kappa_{ib}}{2} b \nonumber \\  & \qquad - \sqrt{\kappa} v_\text{in}e^{-i(\omega_d-\omega_\text{ref})t}.   \label{eq:motion4}
\end{align}

We performed simulations of the system by numerically solving either the differential equations of motion, Eq.~\eqref{eq:motion3} and \eqref{eq:motion4}, or the Lindblad master equation, Eq.~\eqref{eq:master}, using the Python package QuTIP \cite{johansson2012,johansson2013} dedicated to the study of open quantum systems.
In practice, the complex output voltage is measured through heterodyne demodulation of the output signal, which gives its quadrature components $I$ and $Q$. They are defined as $v_\text{out} = I +iQ$.
In order to directly reproduce the experimental data,  
the reference frequency of the rotating frame $\omega_\text{ref}$ has to be set to the frequency of the local oscillator (LO) used to perform the demodulation (see Fig.~\ref{fig2}).
The calculated output voltage is subsequently low-pass filtered,  to account for the finite bandwidth of the antialiasing filter preceding the sampling step in the digitizer. We used a digital Butterworth filter with a cutoff at 90\,MHz.

\subsection{Microwave field oscillation}
\label{section:oscillation}

In order to get an insight to the free evolution of the system, \textit{i.e.}, in absence of a driving field, it is useful to separate the variables $a$ and $b$ in Eqs.~\eqref{eq:motion3} and \eqref{eq:motion4}. We choose $\omega_\text{ref} = \omega_a$. Neglecting the intrinsic losses ($\kappa_{ia}, \kappa_{ib} \ll \kappa $, see Table~\ref{table1}), this yields, for the storage resonator field,
\begin{equation}
 \frac{d^2a}{dt^2} + (\frac{\kappa}{2} + i \Delta) \frac{da}{dt}  + g^2 a = 0. \label{eq:damped_harmonic_oscillator}
\end{equation}

When the resonators are set on resonance, \textit{i.e.}, $\Delta = \omega_b - \omega_a = 0$, the system behaves as a damped harmonic oscillator characterized by the angular frequency $g$ and the damping ratio $\xi = \kappa/(4g)$. 
The field in the storage resonator oscillates in time, as the energy is periodically transferred back and forth between the two resonators. 
It also decays to the transmission line. The decay regime depends on $\xi$. Note that $\xi$ cannot be tuned \textit{in situ} in the experiment; it is a fixed property of each sample.

For an underdamped system ($\xi < 1$), which corresponds to the experiments shown in this article, the decay is slower than the oscillation, and
\begin{equation}
 a(t) = e^{-\frac{\kappa}{4}t} \left(\alpha_1 e^{ig\sqrt{1-\xi^2}t} + \alpha_2 e^{-ig\sqrt{1-\xi^2}t} \right),
\end{equation}
where $\alpha_{1,2}$ are determined by the initial conditions.
The energy, which scales as $|a|^2$, oscillates between the resonators at an angular frequency of $2g\sqrt{1-\xi^2}$. 
In this strong coupling regime of the two resonators \cite{novotny2010}, the effective coupling rate of the storage resonator to the transmission line is $\kappa_\text{eff}=\kappa/2$. It is half of that of the coupling resonator, because, due to the oscillation, the energy is on average only half of the time in the coupling resonator from which it is released to the transmission line.
It is only for a critically damped system ($\xi=1$) that the energy is directly released from the storage resonator to the transmission line at the rate $\kappa/2$ without oscillation. 
For an overdamped system ($\xi > 1$), corresponding to a weak coupling of the two resonators, the decay is more complex, but it is eventually limited by the coupling rate
$\kappa_\text{eff}=\frac{\kappa}{2}\left(1-\sqrt{1-1/\xi^2}\right)$.
For a given $g$, the fastest release of the energy stored in the storage resonator is achieved at the critical damping.

At nonzero detuning, the general solution of Eq.~\eqref{eq:damped_harmonic_oscillator} is
\begin{equation}
 a(t) = \alpha_1 e^{\lambda_+ t} + \alpha_2 e^{\lambda_- t}.   \label{eq:general_solution}
\end{equation}
Again, $\alpha_{1,2}$ are determined from the initial conditions, and 
\begin{equation}
  \lambda_{\pm} = -\frac{1}{2} \left( \frac{1}{2} \kappa + i\Delta \right) \pm \frac{1}{2} \sqrt{ \left( \frac{1}{2} \kappa + i\Delta \right)^2 - 4g^2 }.
\end{equation}
The two resonance modes of the system are involved in the release process, hence the two terms in the general solution. 
The real part of $\lambda_\pm$ indicates the decay of the field whereas their imaginary part corresponds to the field oscillation.

In underdamped systems, the field oscillation between the two resonators becomes faster when the detuning is increased, but only a decreasing fraction of the energy is transferred back and forth,
so that the storage resonator is less and less coupled to the transmission line. The first term of the solution is the most relevant when $\Delta \gg g$. The effective coupling rate can be approximated to 
\begin{equation}
  \kappa_\text{eff} = -2 \operatorname{Re}(\lambda_+) \approx \kappa \frac{g^2}{\Delta^2+2g^2}. \label{eq:kappa_eff_detuning}
\end{equation}

\section{Devices and experimental setup}

\subsection{Samples}

The devices under study are superconducting coplanar waveguide resonators fabricated on the surface of a silicon wafer. The microwave circuit is primarily made of niobium. Only the SQUID, which is located in the middle of the coupling resonator and enables it to be tunable, is made of aluminum. The fabrication process, which has already been described elsewhere \cite{pierre2014}, was specially designed to obtain the longest possible intrinsic lifetime for the resonators, with only two electron-beam lithography steps.

Two samples with two distinct $\xi$ have been studied.
The parameters of the model described in the previous section  corresponding to these samples are shown in Table~\ref{table1}. They have been extracted from the different experiments we performed.
Note that some results on sample II have already been presented \cite{pierre2014}.

\begin{table*}[t]
   \begin{tabular*}{\textwidth}{c @{\extracolsep{\fill}} c c c c c c c c c}
  	\hline \hline
   Sample &  $\omega_a/2\pi$  &  $\omega_b^0 / 2\pi$   &  $\gamma$ & $E_J$  &   $g /2\pi$  &  $\kappa$     & $\xi$ & $\kappa_{ia}$ & $\kappa_{ib}$ \\ \hline    
          &      GHz                  &  GHz           &     $\%$  & meV  &      MHz     & MHz &      & MHz  & MHz  \\ \hline \hline
   I      &   5.416  &  5.844  &  4.8 &  3.1 & 21.2  &  5.0  &  0.0094  &   0.40 &   0.125   \\ \hline
   II     &   5.186  &  5.810  &  8.4 &  2.0 & 18.3  &  280  &  0.61    &   0.054 &  1.8      \\ \hline \hline
   \end{tabular*} 
   \caption{Parameters for the two samples under study: $\omega_a$ resonance frequency of the storage resonator, $\omega_b^0$ bare resonance frequency of the coupling resonator, 
   $\gamma$   inductive ratio for the coupling resonator, $E_J$ maximum Josephson energy of the SQUID, $g$ coupling between the resonators, $\kappa$ coupling rate of the coupling resonator to the transmission line, $\xi$ damping ratio, $\kappa_{ia}$ dissipation rate of the storage resonator, and $\kappa_{ib}$ dissipation rate of the coupling resonator.} 
   \label{table1}
\end{table*}

\subsection{Measurement setup}

\begin{figure}[t!]
\begin{center}
\includegraphics[width=1.0\columnwidth]{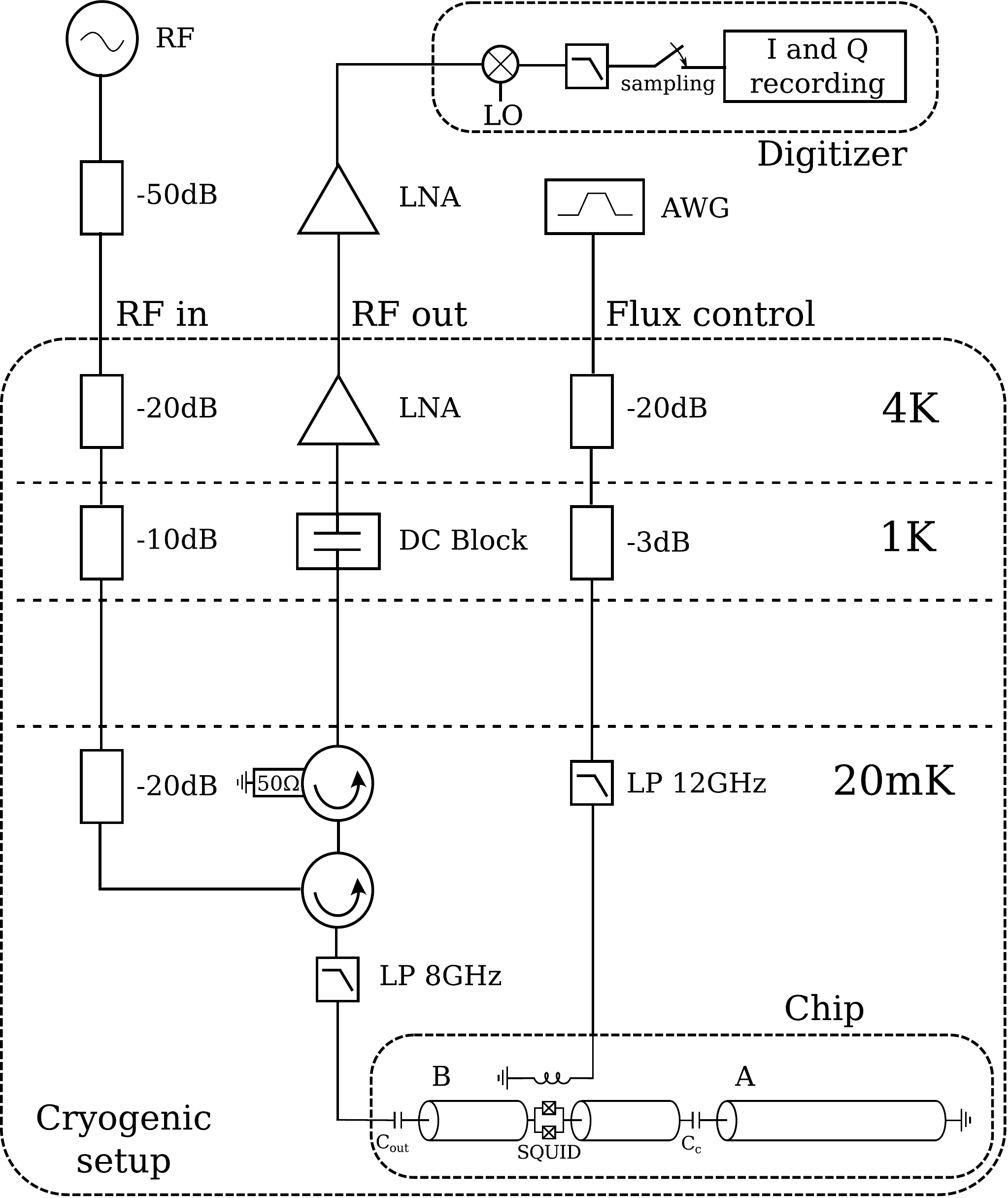}
\caption{Schematics of the cryogenic microwave measurement setup. The chip is located at the coldest stage of a dilution refrigerator equipped with coaxial lines. The input of the device is probed with a reflectometry setup: a microwave signal is routed to the device via a circulator and the reflected signal is amplified with low noise amplifiers, both at 4\,K and room temperature. The output signal is down-converted and then numerically demodulated and sampled with a vector signal analyzer. The control port of the chip is driven with an arbitrary waveform generator (AWG). The input lines are equipped with attenuators and filters in order to prevent room temperature thermal noise from heating the device. The output line is equipped with a circulator acting as an isolator.   
} 
\label{fig2}
\end{center}
\end{figure}

The measurement setup is displayed in Fig.~\ref{fig2}.
The sample is kept below 25\,mK in a dilution refrigerator, wired with coaxial lines. 
The one-port setup at the sample input is transformed into a two-port measurement setup using a circulator to route the microwave signals. This allows one to properly attenuate the input signal, which is necessary for keeping the sample cold and reaching the few-photon level. The reflected signal is amplified with a cold 4-8\,GHz low-noise amplifier from Low Noise Factory.

The input signal, with an RF frequency up to 6\,GHz, can be arbitrarily modulated both in phase and magnitude with a vector signal generator. The output signal undergoes heterodyne demodulation. The resulting quadratures $I$ and $Q$ are sampled at a maximum rate of 200\,Msample/s using a vector signal analyzer.  Both equipments are from Aeroflex and can be synchronously triggered.

The flux in the SQUID loop, which determines the frequency of the coupling resonator, is controlled both by a coil located close to the chip for static biasing and by current pulses applied on-chip. The stability of the system is ensured by magnetic field shielding at low temperature.

\subsection{Characterization of the devices with continuous wave spectroscopy}

The resonance modes of the system are probed by analyzing the reflection coefficient of the devices.
Its magnitude is shown for sample I in Fig.~\ref{fig_spectro}(b) as a function of frequency and the magnetic flux, $\Phi$, in the SQUID loop. 
This experiment is done by measuring the transmission of a signal between the two ports of the setup with a vector network analyzer (VNA, not shown in the experimental setup in Fig.~\ref{fig2}).
The reflection coefficient at the input of the device is obtained from the raw measurement by subtracting the part of the signal which comes from the transmission of the coaxial lines and the microwave components. This background is directly measured for half-integer values of the flux quantum, for which the coupling resonator is detuned away from the measurement band and the storage resonator is strongly undercoupled and therefore not seen in the measurement.

\subsubsection{Resonance mode frequency tuning}

Two resonance modes can be seen for each value of the flux.
The resonance frequency of the coupling resonator evolves periodically with the flux, because of the periodic modulation of the SQUID inductance and critical current. 
It follows that \cite{wallquist2006}
\begin{equation}
  \omega_b(\Phi)  = \frac{\omega_b^0}{1+\frac{\gamma}{\left|\cos \left(\pi \frac{\Phi}{\Phi_0}\right)\right|}},   \label{eq:f_tuning}
\end{equation}
where $\Phi_0 = h/2e$ denotes the flux quantum, $\omega_b^0$ is its bare resonance frequency (\textit{i.e.}, without the SQUID), and $\gamma$ is the inductive participation ratio, defined as the ratio of the SQUID inductance at zero flux over the inductance of the coupling resonator.
The eigenvalues of the Hamiltonian of the system give the resonance frequencies of the two observed resonance modes: 
\begin{equation}
  \omega_\pm(\Phi) = \frac{1}{2}(\omega_a+\omega_b(\Phi)) \pm \sqrt{ g^2 + \left(\frac{\Delta(\Phi)}{2}\right)^2  }.   \label{eq:frequencies}
\end{equation}
They differ from the uncoupled resonance frequencies of the two resonators when they are tuned in resonance, at around $\pm 0.3\,\Phi_0$ in sample I for instance.
A positive detuning is obtained for fluxes around integer number of flux quanta, whereas a negative detuning is obtained around half-integer multiples of flux quanta.

\begin{figure}[t!]
\begin{center}
\includegraphics[width=1.0\columnwidth]{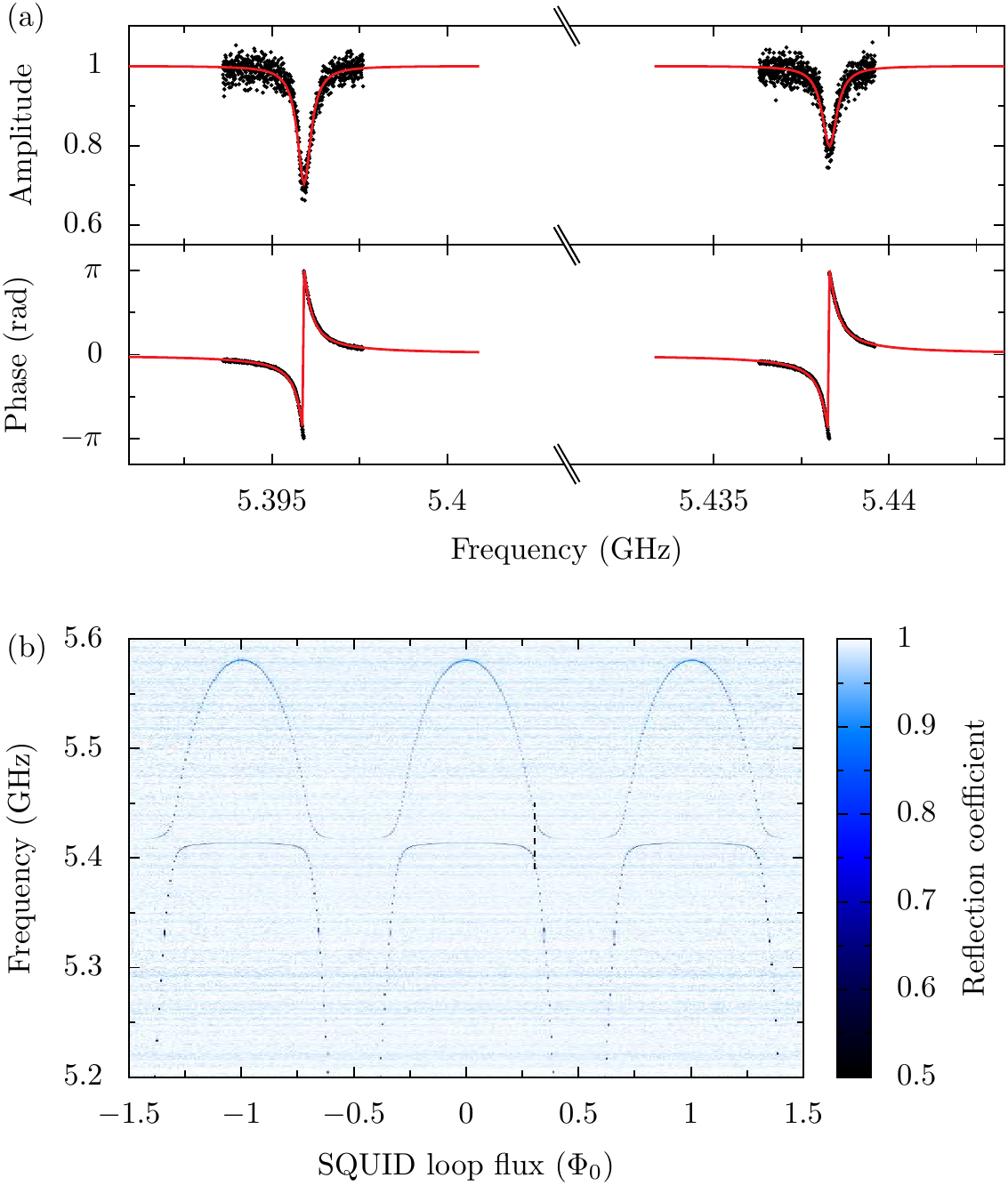}
\caption{Spectroscopy measurements on sample I.
(a) Magnitude and phase of the reflection coefficient for $\Phi=0.3\,\Phi_0$, \textit{i.e.}, when the two resonators are on resonance. Two resonance lines are observed, corresponding to the coupled modes of the two resonators. They are fitted separately with the model given by Eq.~\eqref{eq:gamma_model}.
(b) Magnitude of the reflection coefficient as a function of the frequency and the flux in the SQUID loop. The frequency of the two resonance lines evolve periodically with the flux. The dashed line indicates the cut shown in panel (a).
} 
\label{fig_spectro}
\end{center}
\end{figure}

In underdamped samples, the splitting between the  two modes ($2g$) is larger than their widths, which is of the order of $\kappa/2$ when the two resonators are on resonance. Therefore, the two modes can be treated separately, as if each mode corresponds to a single resonator mode, with a resonance angular frequency $\omega_r$, an effective coupling rate $\kappa_e$, and an effective loss rate $\kappa_i$. 
The equations of motion (Eqs.~\eqref{eq:motion1} and \eqref{eq:motion2}) and the input-output relation (Eq.~\eqref{eq:in-out}) can be adapted for such a single resonator, and their resolution in the frequency domain gives the reflection coefficient
\begin{equation}
 \Gamma = \frac{V_\mathrm{out}}{V_\mathrm{in}} = \frac{\kappa_i - \kappa_e -2i(\omega-\omega_r)}{\kappa_i + \kappa_e -2i(\omega-\omega_r)}.    \label{eq:gamma_model}
\end{equation}
Figure~\ref{fig_spectro}(a) shows the magnitude and the phase of the reflection coefficient at zero detuning. The two resonance modes are fitted separately with Eq.~\eqref{eq:gamma_model}.   
Note that the phase measured with the VNA follows the electrical engineering  rather than the physics convention, thus the substitution $i \leftrightarrow -j$ had to be done in Eq.~\eqref{eq:gamma_model}. Repeating the fitting procedure for every value of the flux allowed us to determine the evolution of the mode resonance frequencies with the flux, as well as their effective coupling rate and loss rate.

The evolution of the extracted resonance frequencies is fitted with Eqs.~\eqref{eq:f_tuning} and \eqref{eq:frequencies}.
This gave us $\omega_a$, $\omega_b^0$, $\gamma$, and $g$ for both samples. These values are shown in Table~\ref{table1}. 
This also gave the evolution of the detuning $\Delta$ with the flux.

\subsubsection{Resonance mode linewidth tuning}

\begin{figure}[t]
\begin{center}
\includegraphics[width=1.0\columnwidth]{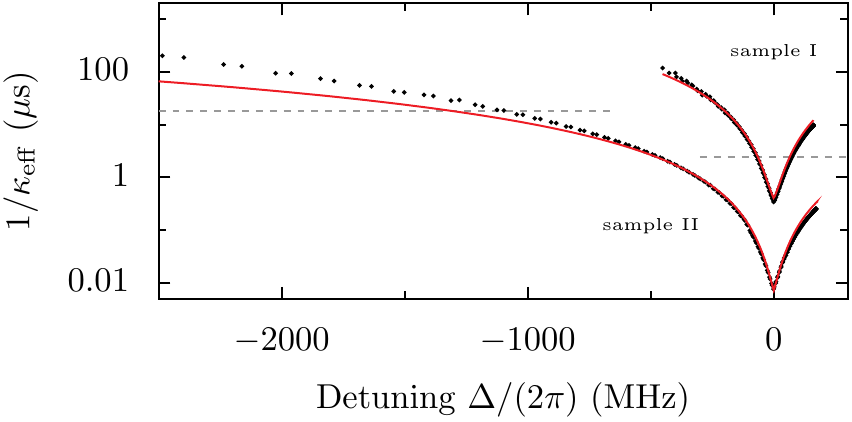}
\caption{Effective coupling time of the storage resonator as a function of the detuning for both samples. The data points come from the fit with Eq.~\eqref{eq:gamma_model} of the reflection coefficient. The plain lines show the model given by Eq.~\eqref{eq:kappa_eff_detuning} without adjustable parameters. The horizontal dashed lines indicate the intrinsic lifetime $1/\kappa_{ia}$ of the storage resonators. In both samples the storage resonator can be set to the over- or under-coupled regimes.
} 
\label{fig4}
\end{center}
\end{figure}

In addition, the fit  of the reflection coefficient gives access to the intrinsic dissipation rate and the coupling rate to the transmission line for each of the resonance modes. 
Figure~\ref{fig4} shows that the effective coupling rate of the storage resonator, plotted against the detuning for every value of the flux, can be varied over several orders of magnitude for both samples. Furthermore, this modulation allows to put the storage resonator in the overcoupled ($\kappa_\text{eff} > \kappa_{ia}$) or undercoupled ($\kappa_\text{eff} < \kappa_{ia}$) regime at will. 
For both samples, the evolution of the effective coupling rate with the detuning is well fitted by Eq.~\eqref{eq:kappa_eff_detuning}, at least when the detuning is not too negative. The model is less accurate at large negative detuning.

Clearly, much larger coupling rates can be achieved with sample II because of a larger $\kappa$, while the same decoupling as in sample I can be obtained, which makes it very suitable for storage and release applications \cite{pierre2014}.

\subsection{Comparison of the behavior of the devices in the resonant coupling regime with time domain spectroscopy}

\begin{figure*}[t]
\begin{center}
\includegraphics[width=1.0\linewidth]{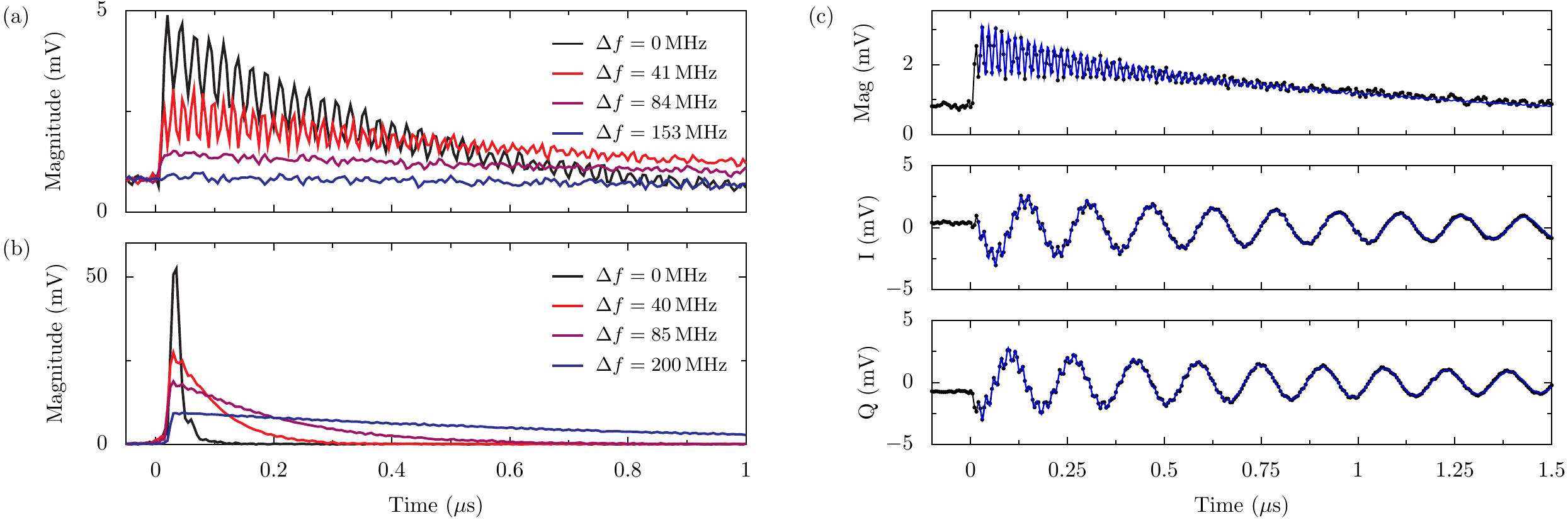}
\caption{Direct release of microwaves from the storage resonator. (a) Output voltage measured for sample I at several detunings. As the detuning is reduced, the decay becomes faster. Beating is observed for low detunings.
(b) Same experiment for sample II.
(c) Magnitude and quadrature signals for sample I for a detuning of $41\,$MHz at $-0.266\,\Phi_0$. Blue line: fit; black line and dots: measurement. The model contains two decaying and oscillating components (see text).
} 
\label{fig6}
\end{center}
\end{figure*}

 The continuous wave spectroscopy measurements shown in Fig.~\ref{fig4} suggest that the release rate of microwaves stored in the storage resonator can be controlled. To study this, we performed a free decay experiment.  In this experiment, the digitizer performing the output signal heterodyne demodulation  is set to the frequency of the storage resonator.

Figure~\ref{fig6} shows a comparison of the behavior of samples I (Fig.~\ref{fig6}(a)) and II (Fig.~\ref{fig6}(b)). Initially, in both cases, the storage resonator is loaded and the coupling resonator is detuned. At $t=0$, the detuning is suddenly reduced. The traces shown corresponds to different final detunings.
Overall, both samples show similar behavior: the smaller the detuning the faster the release of the stored energy is. A noticeable difference is that a much faster release can be achieved with sample II, which simply comes from its larger coupling $\kappa$ between the coupling resonator and the transmission line. This can also be seen in Fig.~\ref{fig4} through the larger range of the effective coupling $\kappa_\text{eff}$. 

The interesting difference lies in the beating that appears for sample I when the detuning is decreased to values of the order of $g$ or smaller.  
This occurs because $\xi \ll 1$ for this sample. Many oscillations of the field take place during the release, thus both coupled modes of the system get populated and decay to the transmission line. The beating of the magnitude of the output signal originates from the interference of their two frequencies.  
The faster beating is observed at zero detuning, where the difference between the frequency of the coupled modes is the smallest. The beating is not observed at too large detunings because one of the two modes is out of the bandwidth of the digitizer. In other words, in the time domain, the beating is faster than the sampling time in this case. 

This effect is better seen on the quadratures of the output signal, which clearly show the superposition of two oscillations at two different frequencies. This can be proven by fitting the quadratures with a two-component model,
\begin{align}
  V_\text{out}(t) &= I(t) + iQ(t) \nonumber \\   
                   &= A_1 e^{-t/\tau_1} e^{-i\left( \omega_1 t + \varphi_1 \right)} + A_2 e^{-t/\tau_2} e^{-i\left( \omega_2 t + \varphi_2 \right)},
\end{align}
as suggested by Eq.~\eqref{eq:general_solution}.
This has been done for all traces, and the excellent agreement is shown in Fig.~\ref{fig6}(c) for $\Delta = 41\,$MHz.

\section{Microwave swapping}

\begin{figure*}[t]
\begin{center}
\includegraphics[width=1.0\linewidth]{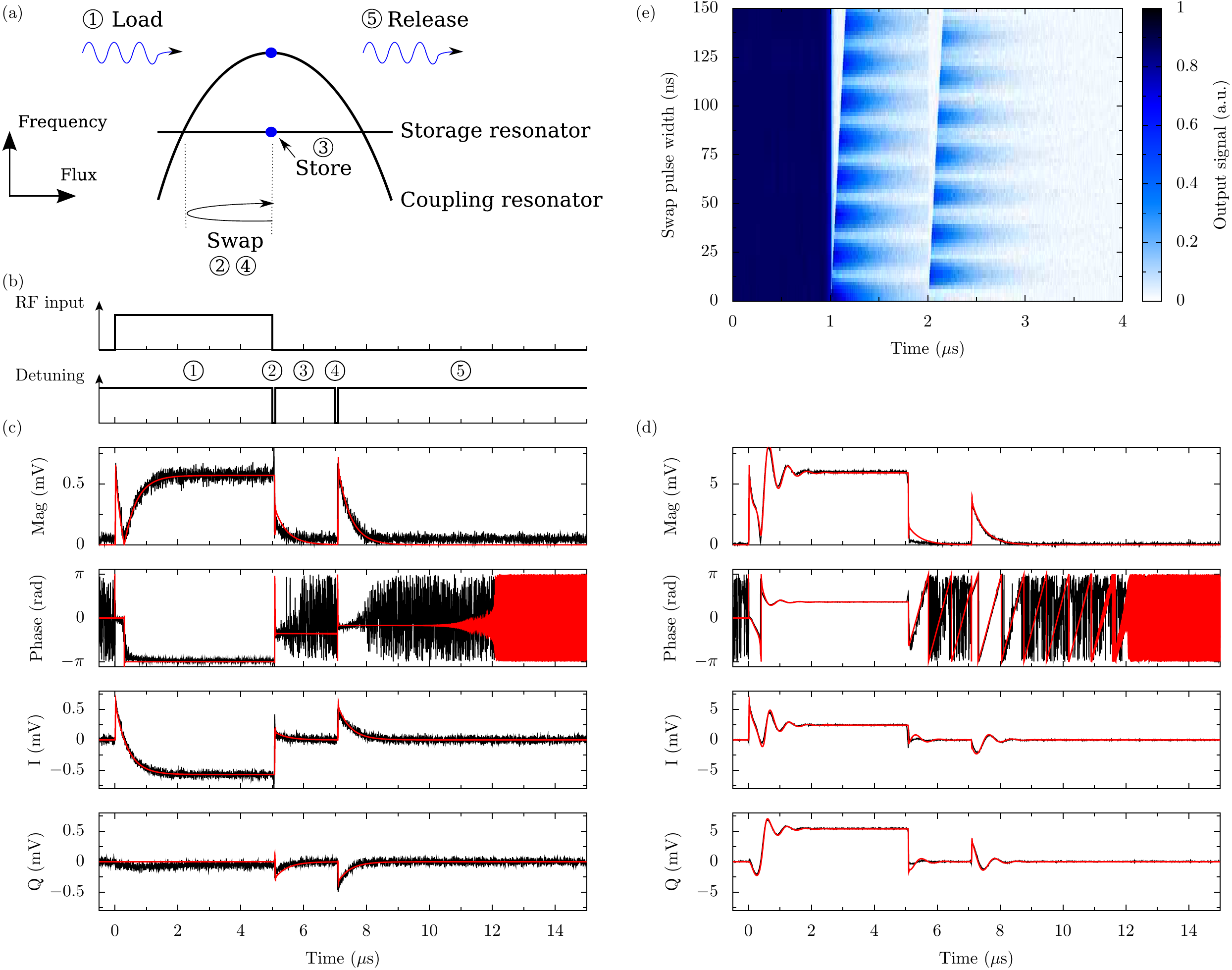}
\caption{Capture and release through field swapping for sample I. (a) Principle of the experiment. A microwave pulse on resonance with the initially detuned (zero flux) coupling resonator is sent to the input port. The detuning is then quickly reduced to zero, leading to a transfer (swapping) of the energy to the storage resonator. After a delay, the swap is repeated, such that the energy is transferred to the coupling resonator from which the microwaves can leak out to the transmission line. 
(b) Measurement sequence showing the RF input and the detuning. Note that the detuning pulses are 10 times shorter than sketched, for clarity reasons.  
(c) Magnitude, phase and quadratures of the measured output voltage for an excitation at the input of the resonator of -139\,dBm. Red lines: simulation. The phase of the measured trace, which has a random reference, has been compensated to match the phase obtained with the simulation.
(d) Same experiment and simulation for an excitation of -119\,dBm, driving the coupling resonator to a nonlinear regime. The signal power is a hundred times higher, hence the reduced noise.
(e) Color map of the output voltage traces for different duration of the swapping pulses. The swap is effective only for certain pulse widths, where the signal is minimum after the first swap pulse, and maximum after the second pulse. For other values, the energy transfer is only partial.}
\label{fig_swap}
\end{center}
\end{figure*}
 
\subsection{Principle}

The two-resonator configuration under consideration allows to tune the effective coupling of the storage resonator to the transmission line. This gives the ability to excite the storage resonator from a resonant incoming wave, at a tunable rate. This can be done only at a moderate detuning between the two resonators. The detuning must be larger than $g$ so that the field builds up only in the storage resonator. It should not be too large, especially when it is negative, as the resonator should be in the overcoupled regime for the energy transfer to be efficient (see Fig.~\ref{fig4}). The range of suitable detuning is nevertheless rather large, in particular for sample II. 
However, in this mode, the coupling rate can only be smaller than $\kappa/2$, which limits the energy transfer speed.

We now present a different scheme, which utilizes the coupling resonator for transferring microwaves to the storage resonator. Whereas this scheme is only possible in samples with $\xi \ll 1$, typically sample I, it allows both faster energy transfer times and incoming waves with various frequencies (detuned from the storage resonator).

The transfer is realized in two steps (see Fig.~\ref{fig_swap}). 
Starting with a large detuning, the coupling resonator is first excited by a resonant incoming microwave pulse with a square envelope. 
Then, when the incoming pulse is turned off, we apply a quick swap operation to transfer the field from the coupling resonator to the storage resonator. This operation is orders of magnitude faster than the decay rate of the coupling resonator, thus a negligible amount of energy is lost instead of being transferred.    

The swap operation is realized by bringing the two resonators on resonance.
The theory developed in Sec.~\ref{section:oscillation} predicts that a periodic energy transfer between the two resonators should occur.
Letting the resonators on resonance for a half integer number of transfer cycles results in a net energy transfer from one resonator to the other.

\subsection{Experimental demonstration}

In practice, a 5-$\mu$s-long square microwave pulse at 5.580\,GHz
is applied at the input of the experimental setup. The power at the input capacitor of sample I is estimated to be -139\,dBm, which corresponds to 17 photons in the pulse. This power has been chosen low enough to ensure a linear response of the coupling resonator. At the end of the pulse, the detuning is quickly decreased, kept at zero for 12\,ns, and then brought back to its initial value. After a delay of 2\,$\mu$s (storage time), the same detuning pulse is repeated. The output signal, measured after heterodyne demodulation and sampling at 200\,MS/s, is shown in Fig.~\ref{fig_swap}(c).  
Its magnitude, its phase, and its quadratures are shown.
In this experiment, the digitizer performing the output signal heterodyne demodulation  is set to the frequency of the coupling resonator, in contrast with the time domain spectroscopy experiment described in the previous section, where it was set to the frequency of the storage resonator. 

The initial rising 
pattern corresponds to the response of the coupling resonator to the input microwave pulse. Since only a rising exponential pulse can be fully absorbed \cite{wenner2014}, a part of the pulse is reflected. Its exact shape depends on the coupling regime ($\kappa/\kappa_{ib}$)
of this resonator. At $t=5$\,$\mu$s, the signal goes quickly towards zero, 
which proves that most of the energy is removed from the coupling resonator after the detuning pulse. The low-amplitude exponential decay corresponds to the release to the transmission line of energy swapped  from the storage resonator, which got excited due to an insufficient detuning during the loading step.
What happens when the detuning pulse is applied cannot be probed with the experiment, first because it is too quick to be seen with the detection sampling rate, and more importantly because both resonators are detuned away from the detection bandwidth during the pulse. When the second detuning pulse is applied, the signal shows a fast rise followed by an exponential decay, which proves that this pulse transfers back some energy stored in the storage resonator to the coupling resonator.

The red line in Fig.~\ref{fig_swap}(c) corresponds to a simulation of the experiment. Equations \eqref{eq:motion3} and \eqref{eq:motion4} are solved using the parameters given in Table~\ref{table1}, which were extracted from the continuous wave measurement. The result is superimposed on the measured traces. Only the amplitude of the traces has manually been adjusted, without any physical significance since the system is in the linear regime. The intrinsic loss rate of the coupling resonator, which is difficult to probe in a spectroscopic experiment because it is much lower than its coupling rate, has also been adjusted to obtain the right shape for the output signal in the loading step. 
The good agreement of the recovered amplitude (after the second detuning pulse) means that the losses, in particular in the storage resonator, are well described by our model. The recovered signal phase is extremely sensitive to the delay between the two pulses, which therefore has been adjusted to obtain the proper distribution of the signal on $I$ and $Q$. 
The good agreement of the simulation gives us access to new information which cannot be probed in the experiment, for instance the resonator populations.

Figure~\ref{fig_swap}(d) shows the same experiment performed with an input power of -119\,dBm, corresponding to 1700 photons in the pulse. This much higher power drives the coupling resonator to a nonlinear regime, of Duffing type, which arises from the intrinsic nonlinearity of the Josephson junctions of the SQUID. The response to the microwave pulse now shows an oscillating pattern. This behavior can be simulated by adding a nonlinear, cubic term $-ib|b|^2$ to Eq.~\eqref{eq:motion4} \cite{krantz2013}. 
The slight shift of the nonlinear resonator frequency has been accounted for in the simulation, which explains
the winding of the phase in the release step.

Figure~\ref{fig_swap}(e) shows that the duration of the detuning pulse must be chosen very precisely. Each horizontal trace corresponds to the output trace of an experiment similar to the one shown in Fig.~\ref{fig_swap}(d), but performed with a variable detuning pulse width, ranging from 0\,ns (no detuning pulse) to 150\,ns. Note that the delay between the two pulses is only 1\,$\mu$s. We observe that the detuning pulse alternatively succeeds and fails to transfer the energy between the resonators. This proves that the mechanism of the energy transfer is a coherent oscillation of the field between the resonators. 
Therefore, when the pulse duration is a full period of this oscillation, the energy simply ends up in the resonator where it was located before the swap pulse. On the other hand, a half-integer number of periods results in swapping the field between the coupling resonator and the storage resonator. The behavior of the output voltage at the second detuning pulse is easily understood: 
when the first pulse keeps the energy into the coupling resonator, it is released to the transmission line directly after, and thus no output 
signal is 
observed when the second pulse is applied.   

This experiment allows us to determine the optimal detuning pulse width, which is $12\pm 1\,$ns. This is in excellent agreement with the predicted value, given by $2\pi/4g \approx 11.8\,$ns.

\section{Conclusion}

We designed, fabricated, and studied superconducting microwave circuits in which we coupled a superconducting resonator to a transmission line through a frequency-tunable resonator. The detuning between these two resonators can be controlled, which enables one to tune the effective coupling of the fixed storage resonator to the transmission line. Moreover, the additional resonance mode introduced by the coupling resonator can get occupied when the two resonators are brought close to resonance. Controlling the detuning therefore also enables the coherent manipulation of microwaves between the two coupled resonators.

The behavior of the system results from the interplay between the oscillation of the field between the coupled resonators and its decay to the transmission line. It depends on a single dimensionless parameter $\xi$ which is the damping ratio of the field oscillation. This parameter can be easily engineered when the circuit is designed by adjusting the coupling capacitances.

This microwave oscillation can be used for catching or releasing microwaves to the transmission line, exploiting the resonant coupling of the two resonators. 
For samples where $\xi \ll 1$, we showed an efficient strategy for storing microwaves in the system, first loading them into the coupling resonator and then swapping them to the storage resonator. 
The latter is done by accurately controlling the detuning in time. This strategy allows to catch or release microwaves within a large range of frequencies since the coupling resonator is tunable, while they are stored at a fixed frequency.

In contrast, the off-resonant coupling of the two resonators allows to release (or catch) microwaves only at the frequency of the storage resonator, but with an effective coupling rate which can be tuned. The tuning range is especially large when $\xi$ is not too small. For $\xi \ll 1$, we showed that the two modes of the system are involved in the release process. This could be useful for creating states of the field with a quantum superposition of frequencies. 

Numerical simulations of the system showed excellent agreement with the experimental data, demonstrating proper modeling of the system. Whereas at low power, close to the single photon level, linear equations could be used, the modeling is also working at higher photon level by introducing a nonlinear term to account for the nonlinear behavior of the coupling resonator originating from the nonlinearity of the SQUID. 

Although we performed experiments with classical signals, it is known that superconducting circuits are suitable for the manipulation of non-classical states of the field \cite{hofheinz2009}. It is likely that the coherent microwaves manipulation which we demonstrated can also be performed with such non-classical states, for instance single photons. The devices described in this article could therefore be implemented as a part of a larger quantum circuit.

\section{Acknowledgments} 

We acknowledge fruitful discussions with our colleagues at the Quantum Device Physics Laboratory and the Applied Quantum Physics Laboratory at Chalmers
University of Technology. 
We also acknowledge financial support from the European Research Council, the European project PROMISCE, the Swedish Research Council, and the Knut and Alice Wallenberg Foundation.

\end{document}